# A microwave photonic prototype for concurrent radar detection and spectrum sensing over an 8 to 40 GHz bandwidth

Taixia Shi [1,†], Dingding Liang [1,†], Lu Wang [2,†], Lin Li [2,*], Shaogang Guo [2], Jiawei Gao [1], Xiaowei Li [1], Chulun Lin [1], Lei Shi [3], Baogang Ding [1], Shiyang Liu [1], Fangyi Yang [1], Chi Jiang [1], and Yang Chen [1,*]

[1] Shanghai Key Laboratory of Multidimensional Information Processing, School of Communication and Electronic Engineering, East China Normal University, Shanghai, 200241, China
[2] Space Optoelectronic Measurement and Perception Lab, Beijing Institute of Control Engineering, Beijing 100190, China
[3] Shanghai Lujie Communication Technology Company Limited, Shanghai, 201613, China
* cast_lilin@163.com
* ychen@ce.ecnu.edu.cn
† These authors contributed equally to this paper

**Abstract:**
In this work, a microwave photonic prototype for concurrent radar detection and spectrum sensing is proposed, designed, built, and investigated. A direct digital synthesizer and an analog electronic circuit are integrated to generate an intermediate frequency (IF) linearly frequency-modulated (LFM) signal with a tunable center frequency from 2.5 to 9.5 GHz and an instantaneous bandwidth of 1 GHz. The IF LFM signal is converted to the optical domain via an intensity modulator and then filtered by a fiber Bragg grating (FBG) to generate only two 2nd-order optical LFM sidebands. In radar detection, the two optical LFM sidebands beat with each other to generate a frequency-and-bandwidth-quadrupled LFM signal, which is used for ranging, radial velocity measurement, and imaging. By changing the center frequency of the IF LFM signal, the radar function can be operated within 8 to 40 GHz. In spectrum sensing, one 2nd-order optical LFM sideband is selected by another FBG, which then works in conjunction with the stimulated Brillouin scattering gain spectrum to map the frequency of the signal under test to time with an instantaneous measurement bandwidth of 2 GHz. By using a frequency shift module to adjust the pump frequency, the frequency measurement range can be adjusted from 0 to 40 GHz. The prototype is comprehensively studied and tested, which is capable of achieving a range resolution of 3.75 cm, a range error of less than ±2 cm, a radial velocity error within ±1 cm/s, delivering clear imaging of multiple small targets, and maintaining a frequency measurement error of less than ±7 MHz and a frequency resolution of better than 20 MHz.

**Keywords:** Radar detection, spectrum sensing, stimulated Brillouin scattering (SBS), microwave photonic, prototype.

## 1. Introduction

Broadband spectrum sensing of radio frequency (RF) signals and high-resolution radar detection are widely applied in cognitive radio systems [1–3], intelligent transportation systems [3, 4], and electronic warfare systems [5–7]. With the continuous and rapid development of these systems, the demand for spectrum sensing and radar functions has gradually shifted from relying on two separate systems to the desire for an integrated system that can achieve both functions, which can greatly facilitate the integration and miniaturization of application platforms.

To fulfill this urgent need, joint radar and spectrum sensing systems have already been investigated in the electrical domain. In joint radar and spectrum sensing systems, the two functions are no longer independent but highly integrated in terms of system architecture and resources. They share some parts of system structures and hardware, and achieve signal multiplexing or sharing on the signal level. However, when aiming to achieve both high-resolution radar detection and broadband spectrum sensing, the integration of radar and spectrum sensing systems via conventional electronic methods encounters significant challenges. Chiefly, these difficulties stem from the "electronic bottleneck", which imposes limitations on the operating bandwidth, especially when striving for simultaneous wide tunability and low system complexity. Microwave photonics [8, 9] uses photonic technology to generate, transmit, and process the RF signal, which can break through the "electronic bottleneck" of the conventional electronic methods and have the distinguished advantage of low transmission loss, good tunability, wide bandwidth, and immunity to electromagnetic interference. In the past few years, microwave photonic radar [10−12], spectrum sensing [13], and integrated multi-function RF systems [14−15] have garnered extensive research attention from both academic and industrial communities.

Microwave photonics possesses the capability to enhance the key performance indicators of individual radar or spectrum sensing systems. With the assistance of microwave photonics, the challenges of generating and receiving ultra-wideband wide-range frequency-tunable radar signals are readily addressed, which greatly improves the flexibility and anti-interference ability of radar systems [16]. The classical photonics-assisted radar signal generation methods can be implemented by photonic frequency conversion [17, 18], photonic frequency multiplication [19, 20], optical injection [21], photonic digital-to-analog converter [22], and acoustic-optic frequency shifting loops [23]. For wideband radar signal reception, microwave photonic channelization [24], photonic de-chirping [17−20], and photonic compressed sensing [25] can be applied. Microwave photonics can also be applied to spectrum sensing systems, where compared to conventional electronics solutions, its main contribution is the ability to quickly acquire frequency information within an ultra-wide frequency measurement range. Although its frequency accuracy may be inferior to electronics solutions, the rapid acquisition of frequency information is crucial in certain application scenarios. To acquire the frequency information of the signal under test (SUT), microwave frequency measurement [26−30] and time-frequency analysis [31−35] can be employed. The basic principle is to map the frequency information of the SUT to other optical or electrical parameters that are easier to measure, which can be implemented via frequency-to-power mapping (FTPM) [26, 27], frequency-to-space mapping (FTSM) [27, 28], and frequency-to-time mapping (FTTM) [29−35]. In general, the method based on FTPM is only applicable for single-frequency measurement unless it is combined with other methods [27]. The method based on FTPM should also be used in conjunction with other methods [27], otherwise, the measurement resolution is poor unless there is a large number of channels at the cost of great complexity [28]. The method based on FTTM can achieve multi-frequency measurement with suitable system complexity and good measurement accuracy. In addition, this kind of method has good scalability and can be further extended to analyze the two-dimensional time-frequency parameter [31−35].

In recent years, microwave photonic systems that deeply integrate radar and spectrum sensing functions have been studied. It is hoped that with an integrated system, both these two RF functions can be achieved simultaneously by sharing part of the signal and the hardware. In these proposed schemes [36−39], different photonic methods are employed to generate the linearly frequency-modulated (LFM) signals and to implement the de-chirping operation for radar function, whereas the SUT is optically converted to an electrical frequency-sweep signal and its frequency is mapped to the time domain using a narrow-bandwidth electrical bandpass filter (EBPF) for spectrum sensing function. In [36, 37], tunable frequency-multiplied LFM signal generation, radar echo de-chirping, and modulation and measurement of the SUT are realized through optical sideband manipulation. However, the system structure is relatively complex, requiring a complex modulator structure and optical bandpass filter (OBPF) for sideband generation and selection, which limits the tunability of the system to a certain extent. In [38], the two functions do not share the same frequency-sweep source, which reduces the interdependence of the two functions but also creates some redundancy in the system. In [39], optical injection of a semiconductor laser is employed to generate the frequency-sweep source for both radar and spectrum sensing functions, which reduces the cost and complexity of the system. However, due to the limited performance of the frequency-sweep optical signal generated by optical injection, the frequency measurement error and resolution in [39] are the worst among these schemes. Furthermore, because all these schemes in [36−39] realize the FTTM in the electrical domain, in the spectrum sensing function, high-speed photodetectors (PDs) are needed before the EBPF to generate the electrical frequency-sweep signal carrying the frequency information of the SUT, which increases the cost of the system. Moreover, all these proposed schemes are only studied and demonstrated in an experiment, and each of them has realized only a part of the many functions of the joint radar and spectrum sensing system, including ranging, radial velocity measurement, imaging, frequency measurement, and time-frequency analysis. In addition, the operating frequency range and tunability of the systems also need to be further improved.

In this work, to the best of our knowledge, the first microwave photonic prototype for concurrent radar detection and spectrum sensing is proposed, designed, built, and investigated. An IF LFM signal is generated by a direct digital synthesizer (DDS) working in conjunction with an analog electronic circuit. The IF LFM signal has a frequency-sweep bandwidth of 1 GHz and is converted to the optical domain at an intensity modulator (IM). The optical carrier of the optical signal from the IM is filtered out by a fiber Bragg grating (FBG), and only the ±2nd-order LFM optical sidebands are left. The ±2nd-order LFM optical sidebands beat with each other to generate the frequency-and-bandwidth-quadrupled LFM signal for radar applications, while one of the ±2nd-order LFM optical sidebands is selected by another FBG and modulated by the SUT. The optical signal carrying the SUT then works in conjunction with a stimulated Brillouin scattering (SBS) gain spectrum to map the frequency of the SUT to the time domain for spectrum sensing function. By tuning the frequency of the IF LFM signal, the radar function of the prototype can be operated with a 4-GHz instantaneous bandwidth in a frequency range from 8 to 40 GHz. The spectrum sensing range of the prototype can be extended from 0 to 40 GHz by adjusting the pump wave frequency and the instantaneous analysis bandwidth is 2 GHz. The prototype is comprehensively studied and tested, which is capable of achieving a range resolution of 3.75 cm, a range error of less than ±2 cm, a radial velocity error within ±1 cm/s, delivering clear imaging of multiple small targets, and maintaining a frequency measurement error of less than ±7 MHz and a frequency resolution of better than 20 MHz.

## 2. Principle and setup of the prototype

2.1 System architecture

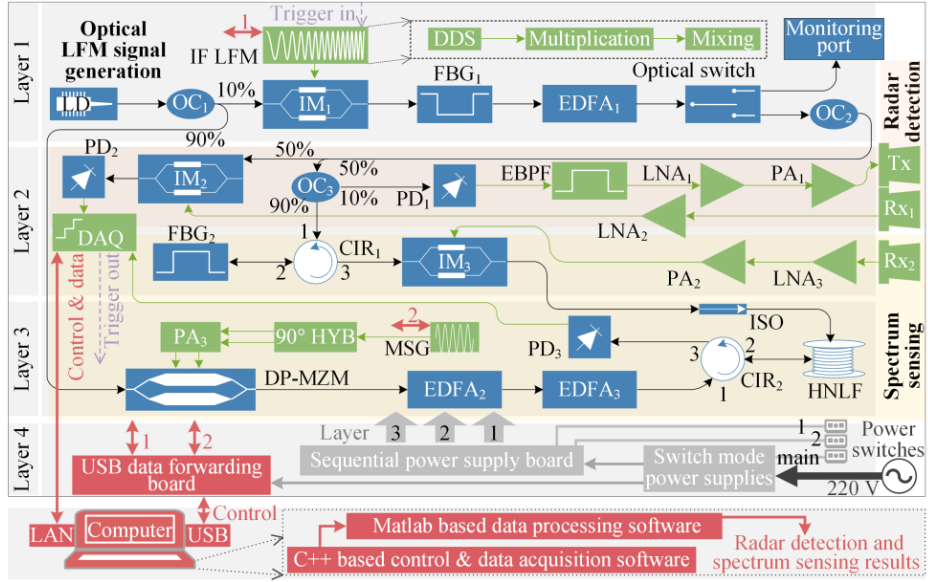

Fig. 1. Schematic diagram of the photonics-assisted concurrent radar detection and spectrum sensing prototype. LD, laser diode; OC, optical coupler, IM, intensity modulator; IF, intermediate frequency; LFM, linearly frequency-modulated signal; DDS, direct digital synthesizer; FBG, fiber Bragg grating; EDFA, erbium-doped fiber amplifier; PD, photodetector; EBPF, electrical bandpass filter; LNA, low-noise amplifier; PA, power amplifier; Tx, transmitting antenna; Rx, receiving antenna; DAQ, data acquisition board; CIR, circulator; ISO, isolator; HNLF, highly nonlinear fiber; DP-MZM, dual-parallel Mach–Zehnder modulator; MSG, microwave signal generator; 90° HYB, 90° hybrid coupler; USB, universal serial bus; LAN, local area network.

The system architecture of the microwave photonic prototype for concurrent radar detection and spectrum sensing is depicted in Fig. 1. The prototype directly uses a 220 V AC power supply. We have developed a control and data acquisition software using C++, as well as a data processing software using Matlab, for it. Its control, data transmission, and processing are achieved by using a computer. According to the functions implemented, the prototype is divided into 4 layers: Layer 1 is used to generate the optical LFM signal; Layer 2 is used for radar signal generation and detection of the radar function, the SUT reception of the spectrum sensing function, and the data acquisition function; Layer 3 is used for frequency information identification of the spectrum sensing function; Layer 4 integrates the power supply and command receiving and forwarding function. The rear panel of the prototype provides three interfaces, including a main power switch, a grounding port, and an optical monitoring port for observing the optical spectrum from an erbium-doped fiber amplifier (EDFA$_1$) via an optical switch in layer 1. The front panel of the prototype has five interfaces, including two universal serial bus (USB) ports, one local area network (LAN) port, and two power switches. One USB port is reserved for future use, while the other serves to establish a connection between the computer and the IF LFM signal generator located on layer 1, as well as the MSG situated on layer 3, through a USB data forwarding board on layer 4. The LAN port is used to connect the computer and the data acquisition board (DAQ) on layer 2 for data transmission and DAQ control. The two power switches are connected to the sequential power supply board on layer 4 to sequentially power up all the components that need to be powered in the prototype. All control commands are set and generated using the control and data acquisition software and then forwarded to the IF LFM signal generator and MSG via the USB port and to the DAQ via the LAN port. The data acquired by the DAQ is transmitted back to the computer via the LAN port, and stored locally as a file, which is further processed in the data processing software to get the radar detection and spectrum sensing results. In the process of data acquisition, the DAQ will output a trigger signal to the IF LFM signal generator for synchronization. Therefore, the reference signal used for absolute frequency positioning [34] can be removed from the prototype.

## 2.2 Optical LFM signal generation

First, the optical LFM signal generation in the prototype is explained. In layer 1 of the prototype, a narrow-linewidth laser diode (LD, Photonteck PHX-C-F-M-C34-13-6-1-0-0) is used to generate a continuous-wave light wave with a power of 13 dBm and a wavelength of 1549.919 nm, and the corresponding frequency is denoted as $f_0$. The optical signal from the LD is divided into two parts by an optical coupler ($OC_1$). The 10% output of $OC_1$ is modulated by an IF LFM signal at $IM_1$ (Fujitsu FTM7938EZ). The IF LFM signal has a power of around 15 dBm, a period $T$ of 1.4876 ms, a bandwidth of 1 GHz, and an adjustable center frequency $f_C$ from 2.5 to 9.5 GHz. The IF LFM signal is expressed as $E_{IF}(t) = E_0\cos(2\pi f_C t + \pi k t^2)$, where $E_0$ and $k$ are its amplitude and chirp rate, respectively. $IM_1$ is biased and stabilized at the maximum transmission point (MATP) using a modulator bias controller ($MBC_1$, PlugTech MBC-NULL-03). Under the small signal modulation condition, only the optical carrier and ±2nd-order optical sidebands are taken into account. The optical signal output from $IM_1$ is injected into $FBG_1$. The temperature of $FBG_1$ is controlled by a temperature controller with a temperature stability of 0.5 °C. The transmission spectrum of $FBG_1$ is used as a narrow-bandwidth optical bandstop filter (OBSF) to suppress the optical carrier and leave only the ±2nd-order optical sidebands of the IF LFM signal. $EDFA_1$ (Max-Ray EDFA-C-PA-45-SM-M) is used to boost the power of ±2nd-order optical sidebands with frequencies of $f_0 \pm (2f_C + 2kt)$. The amplified ±2nd-order optical sidebands serve as the optical LFM signal, which is used for the following radar and spectrum sensing functions. In the radar function, the optical LFM signal is used for radar signal generation and also used as an optical reference signal for radar de-chirping. In the spectrum sensing function, the optical LFM signal is used as a frequency-sweep optical signal for further signal modulation and FTTM.

## 2.3 Radar detection

The radar detection function in the prototype is realized using the optical LFM signal generated in layer 1. The optical LFM signal from layer 1 is sent to $OC_2$ via an optical switch. Note that the other output port of the optical switch is connected to the rear panel for monitoring. When power switch 1 on the front panel is on and power switch 2 on the front panel is off, the input optical signal is sent to the monitor port on the rear panel. As power switch 2 is on, the switch is supplied with a 5 V DC, so the optical LFM signal from layer 1 is sent to $OC_2$. The two outputs of $OC_2$ are sent to $OC_3$ and $IM_2$ (Fujitsu FTM7938EZ) on layer 2 of the prototype, respectively. The 10% output of $OC_3$ is injected into $PD_1$ (Coherent XPDV2120R), in which the two ±2nd-order optical sidebands of the optical LFM signal beat with each other to generate a frequency-and-bandwidth-quadruple LFM signal, i.e., the radar signal. The bandwidth of the generated radar signal is 4 GHz, whereas its center frequency is adjustable from 10 to 38 GHz. Subsequently, the radar signal is filtered by the EBPF (Talent Microwave TLHF-8G-40G-X), amplified by a low-noise amplifier ($LNA_1$, Talent Microwave TLLA1G40G-40-45) and a power amplifier ($PA_1$, Connphy Microwave CMP-0.1G40G-3020-K), and finally radiated to the free space through a transmitting antenna (TX). The radar echo from the targets is collected by a receiving antenna ($RX_1$), amplified by $LNA_2$ (Talent Microwave, TLLA1G40G-40-45), and then applied to the RF port of $IM_2$. $IM_2$ is biased and stabilized at the quadrature transmission point by $MBC_2$ (PlugTech MBC-MZM-01) to implement the mixing of the optical reference signal and the radar echo. After mixing in $IM_2$, the optical signal from $IM_2$ is detected in $PD_2$ (Nortel PP-10G) to realize the radar de-chirping. The low-frequency radar de-chirped signal from $PD_2$ is sampled by the DAQ. The digitized radar de-chirped signal is then sent to the computer where it is further processed.

The radar de-chirped signals with different durations are utilized to extract the range, radial velocity, and inverse synthetic aperture radar (ISAR) imaging. The target range is accurately determined by applying a fast Fourier transform (FFT) to a single-period radar de-chirped signal. Meanwhile, the radial velocity of the target is calculated by measuring the range variation over a 0.5-s interval. For ISAR imaging, the range-Doppler algorithm is employed [17], which necessitates a sampling time of 1.5 s.

2.4 Spectrum sensing

The spectrum sensing function of the prototype is also realized using the optical LFM signal generated in layer 1. The optical LFM signal from $OC_2$ in layer 1 is equally split by $OC_2$, and one output of $OC_2$ is split by $OC_3$ in layer 2. The 10% output of $OC_3$ is used for the radar function discussed above, while the 90% output of $OC_3$ is sent to $FBG_2$ via a circulator ($CIR_1$). $FBG_2$ is also thermostatically controlled by another temperature controller. The reflection spectrum of $FBG_2$ functions as an OBPF. The center wavelength and 3-dB bandwidth of $FBG_2$ are 1550.015 nm and 18 GHz. After being reflected by $FBG_2$, the –2nd-order optical sideband with a frequency of $f_0 - (2f_C + 2kt)$ is selected and then sent to $IM_3$ (Fujitsu FTM7938EZ). $IM_3$ is biased and stabilized at the minimum transmission point by $MBC_3$ (PlugTech MBC-NULL-03). The SUT collected by another receiving antenna ($RX_2$) is amplified by $LNA_3$ (Talent Microwave TLLA1G40G-40-45) and $PA_2$ (Centellax, OA4SMM5) and applied to the RF port of $IM_3$. After modulation, the SUT is converted to a series of frequency-sweep optical sidebands in the optical domain. The output of $IM_3$ is used as a probe wave and sent to layer 3, in which it is connected to a 200-m highly nonlinear fiber (HNLF, YOFC NL1016-B) via an optical isolator (ISO).

The 90% output of $OC_1$ in layer 1 is also distributed to layer 3, in which it is injected into a dual-parallel Mach–Zehnder modulator (DP-MZM, Fujitsu FTM7961EX). A single-tone RF signal with an adjustable frequency $f_s$ from 0.1 to 40 GHz is generated by an MSG and applied to the RF ports of the DP-MZM via a 90° hybrid coupler (90° HYB, Talent Microwave TBG-20400-3k-90) and $PA_3$ (Centellax OA4SMM4). The DP-MZM is biased as a carrier-suppressed single-sideband (CS-SSB) modulator using $MBC_4$ (PlugTech MBC-IQ-03), shifting the optical carrier applied to the DP-MZM according to the frequency of the RF signal. Then, the optical signal from the DP-MZM is used as a pump wave and reversely sent to the HNLF through $EDFA_2$ (Max-Ray, EDFA-C-PA-45-SM-M), $EDFA_3$ (Max-Ray, EYDFA-C-HP-BA-35-SM-M), and $CIR_2$ to generate an SBS gain. The SBS gain spectrum acts as a narrow bandwidth optical filter besides providing a beneficial gain. Since different frequency-sweep optical sidebands generated by different SUT frequencies in the probe wave are filtered by the SBS gain spectrum at different times, optical pulses will be generated at different times in a single sweep period according to the SUT frequency. Therefore, the frequency of the SUT is mapped to the time domain when the optical pulse appears in a sweep period via FTTM. Finally, the optical pulses from port 3 of $CIR_2$ are injected into $PD_3$ (Nortel PP-10G) and converted to low-speed electrical pulses. The electrical pulses from $PD_3$ are sampled by the DAQ and further processed in the computer for frequency measurement and time-frequency analysis.

The instantaneous frequency analysis bandwidth is 2 GHz, which is determined by the frequency-sweep bandwidth of the optical LFM signal. Nevertheless, by changing the frequency $f_s$ applied to the DP-MZM, i.e., changing the pump wave frequency by $f_a = \pm f_s$ according to the direction of the frequency shifting, the 2-GHz instantaneous frequency analysis bandwidth can be located at any location from 0 to 40 GHz. When the pump frequency is not shifted, i.e., $f_a = 0$, the center frequency of the SBS gain is $f_0 - f_{SBS}$, where $f_{SBS}$ is the Brillouin frequency shift and is 9.4 GHz using the 200-m HNLF. When the pump frequency is shifted by $f_a = \pm f_s$, the center frequency of the SBS gain is also shifted to $f_0 + f_a - f_{SBS}$. Generally, the frequency measurement range is from $f_a - f_{SBS} + 2f_C - kT$ to $f_a - f_{SBS} + 2f_C + kT$. Here, $f_a - f_{SBS} + 2f_C - kT$ should be greater than 0. In this work, if the nonlinear medium is not changed and the sweep bandwidth of the IF LFM signal is also kept unchanged, the instantaneous frequency analysis range is from $f_a + 2f_C - 10.4$ GHz to $f_a + 2f_C - 8.4$ GHz.

## 3. Test results of the prototype and discussion

### 3.1 Generation of the optical LFM signal and radar signal

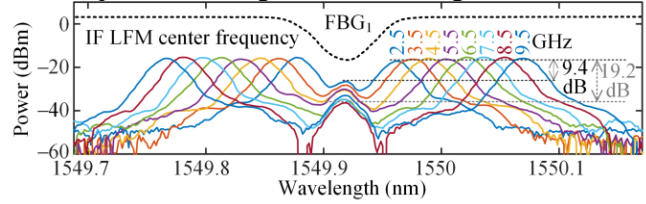

Fig. 2. Optical spectra of the optical LFM signal under different center frequencies of the IF LFM signal. The black dotted line shows the transmission spectrum of FBG$_1$.

Layer 1 of the prototype is primarily used to generate the optical LFM signal for the radar and spectrum sensing functions of the prototype. First, the generation of the optical LFM signal and radar signal is demonstrated. The modulated optical signal from IM$_1$ is filtered by FBG$_1$ to suppress the optical carrier. FBG$_1$ has a center wavelength of 1549.919 nm and a 3-dB bandwidth of 10 GHz, and its transmission spectrum is equivalent to an OBSF, as indicated by the black dotted line in Fig. 2. The optical LFM signal from FBG$_1$ with its carrier suppressed is amplified and then monitored by an optical spectrum analyzer (OSA, ANDO AQ6317B) via the monitoring port. Figure 2 shows the optical spectra of the optical LFM signal under different center frequencies of the IF LFM signal from 2.5 to 9.5 GHz. As can be seen, the ±2nd-order optical sidebands are dominant with a suppressed optical carrier. The peak power of the LFM sidebands is almost the same because the IF LFM signal power is pre-set and adjusted at different frequencies according to the response of the modulator. However, the carrier suppression ratios for different IF LFM signals are slightly different, which changes from 9.4 to 19.2 dB. The main reason for the different carrier suppression ratios is that the power of the IF LFM signal is different at different frequencies, resulting in the carrier power of the FBG$_1$ input is also different.

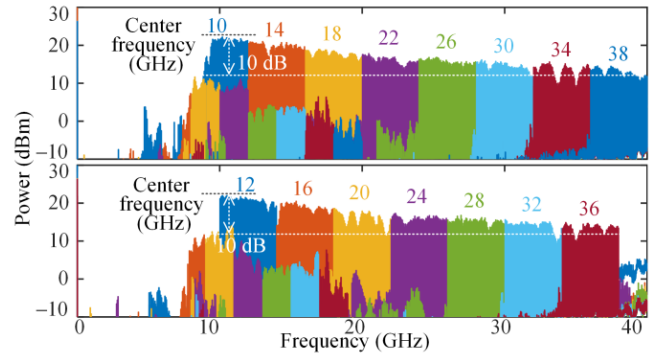

Fig. 3. Electrical spectra of the generated 4-GHz bandwidth LFM radar signal ranging from 8 to 40 GHz.

The optical LFM signal generated above is detected in the high-speed PD$_1$ in layer 2 to generate the LFM radar signal for the prototype. After being filtered by the EBPF and amplified by LNA$_1$ and PA$_1$, the electrical spectra of the LFM radar signal are measured by an electrical spectrum analyzer (ESA, R&S FSP-40) with a resolution bandwidth of 3 MHz, a video bandwidth of 10 MHz, and a sweep time of 1 s, as shown in Fig. 3. When the center frequency of the IF LFM signal is adjusted from 2.5 to 9.5 GHz with a step of 0.5 GHz, the center frequency of the generated LFM radar signal changes from 10 to 38 GHz with a step of 2 GHz and the bandwidth of the LFM radar signal is quadrupled to 4 GHz.

As can be seen from Fig. 3, as the frequency of the LFM radar signal increases, its power decreases, which is mainly attributed to the frequency response of $LNA_1$, $PA_1$, and $PD_1$. In building the prototype, we do not compensate for the power inconsistencies for two main reasons: 1) It is very difficult to further increase the power of the IF LFM signal at high frequencies, and 2) We do not have a programmable automatic gain amplifier at such a large bandwidth. Furthermore, a certain gain unevenness is observed in the electrical spectrum, which is largely introduced by standing waves in the measurement process. In addition, it can be clearly seen that the generated LFM radar signal centered at 10 GHz has lower power from 8 to 9 GHz, which is mainly caused by the 10-GHz bandwidth of $FBG_1$. In addition to filtering out the optical carrier, $FBG_1$ also suppresses part of the optical sidebands to a certain extent when the sidebands are close to the carrier. This issue can be solved by using an FBG with a smaller 3-dB bandwidth instead of $FBG_1$.

3.2 Ranging and radial velocity measurement

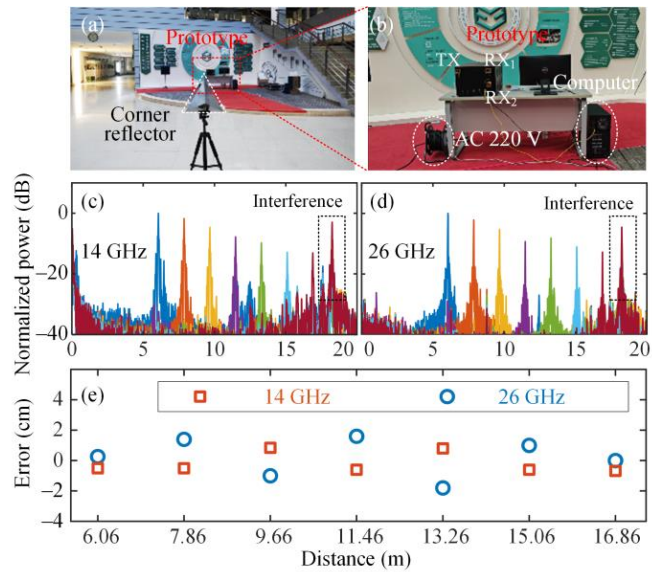

Fig. 4. (a) Photograph of the ranging experimental setup. (b) A zoomed-in view of the red-dotted rectangular area in (a). Ranging results of the corner reflector at different distances when the center frequencies of the LFM radar signal are (c) 14 GHz and (d) 26 GHz, respectively. (e) Ranging errors at different distances and center frequencies of the LFM radar signal.

Figure 4(a) shows the setup of the ranging experiment using the prototype and a zoomed-in view of the prototype is shown in Fig. 4(b). In the experiment, a corner reflector is placed in front of the prototype as a static target. By changing the distance between the corner reflector and the antenna pair at an interval of 1.8 m, the range of the corner reflector is successively measured when the center frequencies of the LFM radar signal are 14 and 26 GHz, respectively. Ranging results of the corner reflector are demonstrated in Figs. 4(c)-4(e). Due to the photonic de-chirping, the echo reflected by the corner reflector is compressed in the frequency domain, with the spectra of the de-chirped signal shown in Figs. 4(c) and 4(d), respectively. As can be seen, with the increase of the distance, the peak in the frequency domain also increases. The target range $L$ is determined via the peak frequency by $L = cf_p/(8k)$ [19], where $c$ is the velocity of the electromagnetic wave, $f_p$ is the peak frequency. It is also noted that a fixed peak at 18.24 m always exists, which is mainly caused by the reflection of a column facing the prototype. The range errors are shown in Fig. 4(e), which is no more than ±2 cm.

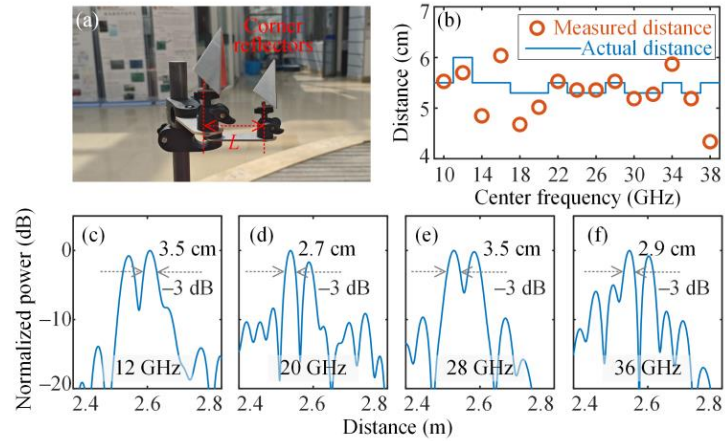

Fig. 5. (a) Photograph of the two smaller corner reflectors placed close to each other. (b) Measured distance between two smaller corner reflectors at different center frequencies. (c)-(f) Zoomed-in view of the de-chirped signal spectra at center frequencies of 12, 20, 28, and 36 GHz.

To further verify the ranging performance of the prototype, two smaller corner reflectors close to each other are employed as the ranging targets, as depicted in Fig. 5(a). The instantaneous bandwidth of the LFM radar signal is fixed at 4 GHz, while the center frequency of the signal is adjusted from 10 to 38 GHz with a step of 2 GHz. The distance between the two corner reflectors is adjusted to the minimum distance that can be fully and easily distinguished when the prototype works at different frequency bands. As shown in Fig. 5(b), the distance between the two smaller corner reflectors is set to between 5.3 and 6 cm over the whole operating bandwidth. The circles represent the measured distance between the two corner reflectors while the solid line represents the actual distance. The errors are less than 1.5 cm. Figures 5(c)-5(f) show the zoomed-in view of the de-chirped signal spectra when the center frequencies of the radar LFM signal are 12, 20, 28, and 36 GHz. The corresponding distances between the two corner reflectors are 6, 5.3, 5.5, and 5.3 cm, respectively. It can be seen that two corner reflectors are easily distinguished. It is actually possible to distinguish even closer distance and only the cases that can be fully and easily distinguished are given in the experiment. Theoretically, the range resolution of the prototype with an operating bandwidth of 4 GHz is 3.75 cm. As shown in Figs. 5(c)-5(f), the full width at half maximum (FWHM) of the peaks is from 2.9 to 3.5 cm, which is close to the theoretical value.

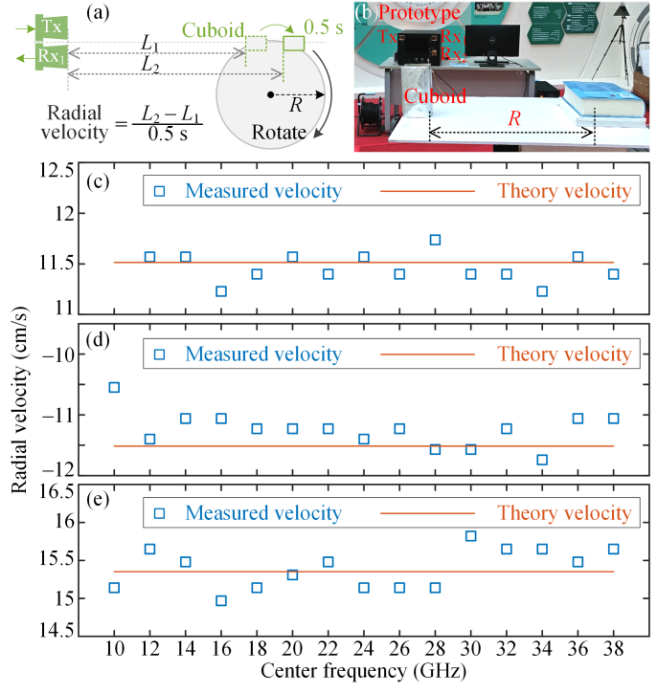

Fig. 6. (a) Schematic diagram of the radial velocity measurement. (b) Photograph of the radial velocity measurement setup. Radial velocity measurement results when the target rotates (c) clockwise and (d) counterclockwise with a rotation radius of 45 cm. (e) Radial velocity measurement results when the target rotates clockwise with a rotation radius of 60 cm.

Then, the radial velocity measurement using the prototype is verified. Figure 6(a) shows the schematic diagram of the radial velocity measurement. A cuboid is positioned on the turntable, maintaining a distance of $R$ from the center of the rotating platform. The turntable rotates in a horizontal plane with an angular velocity of $\omega$, either clockwise or counterclockwise. When the velocity direction of the target is consistent with the radar line of sight, as shown in Fig. 6(a), a measurement of the target's range at two extremely proximate instances in time is sufficient, from which the radial velocity of the target can be derived based on the rate of change in range. The reason for conducting radial velocity measurement through the aforementioned method is that we lack the equipment that can support the uniform linear motion of a target.

In the experiment, the angular velocity of the turntable is $\omega = 0.256$ rad/s, and the time interval for two range measurements is set to 0.5 s. The time interval cannot be too small to reduce the radial velocity error caused by ranging errors; neither can it be too large, as an overly large interval would result in a significant deviation in the target's velocity direction between the two range measurements, which would also increase the radial velocity measurement error. Under these circumstances, the radial velocity of the cuboid can be obtained by $(L_2–L_1)/(0.5\text{ s})$, while the theory value of the radial velocity is obtained by $\omega \times R$. Figure 6(b) shows the radial velocity measurement setup. When the cuboid rotates clockwise or counterclockwise with a rotation radius of 45 cm, the radial velocity of the cuboid is measured and shown in Figs. 6(c) and 6(d). In this case, the theoretical radial velocities of the target when rotated clockwise and counterclockwise are 11.5 and −11.5 cm/s, respectively. When the rotation radius of the cuboid is changed to 60 cm, the radial velocity measurement results are depicted in Fig. 6(e). As can be seen from Figs. 6(c)-6(e), the measured radial velocity fluctuates around the theoretical value, and the errors are all less than ±1 cm/s.

3.3 Small-target ISAR imaging

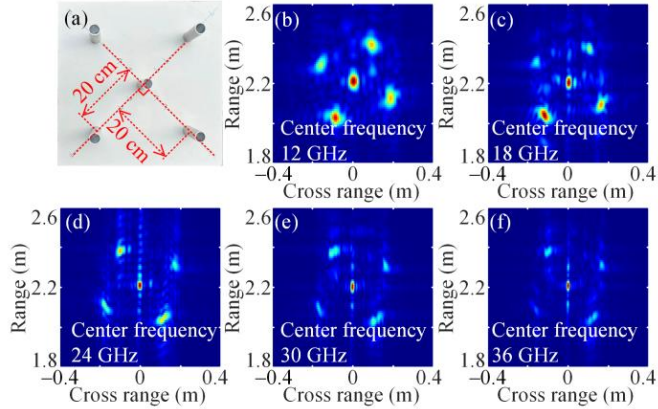

Fig. 7. (a) Photograph of five cylinders with an outer diameter of 3 cm for ISAR imaging. (b)-(f) ISAR imaging results obtained using the prototype operating at different center frequencies.

Subsequently, the capability of ISAR imaging of the prototype is demonstrated. Five cylinders made of aluminum with an outer diameter of 3 cm are selected as the targets and they are placed on the turntable as shown in Fig. 7(a). The angular velocity of the turntable is also $\omega = 0.256$ rad/s. Along the radar line of sight, the distance between the center of the turntable and the antenna pair is 2.22 m. The accumulation time in ISAR imaging is 1.5 s. Figures 7(b)-7(f) show the ISAR imaging results of the five cylinders. The center frequencies of the radar LFM signal are 12, 18, 24, 30, and 36 GHz. The five targets are easy to distinguish and identify. However, as the frequency increases, due to the lower signal power and greater transmission loss, the target is weaker and weaker in the images.

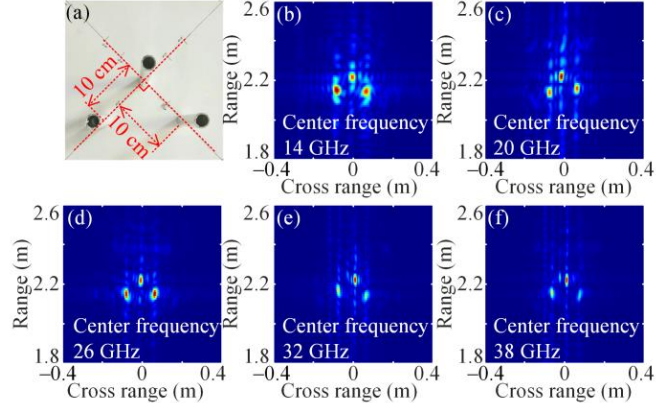

Fig. 8. (a) Photograph of three cylinders with an outer diameter of 2.5 cm for ISAR imaging. (b)-(f) ISAR imaging results obtained using the prototype operating at different center frequencies.

Then, the distance between adjacent cylinders is further reduced to show the ISAR imaging performance. Besides, the center frequency of the LFM radar signal is also changed to show the imaging capability of the prototype at other frequencies. Another three cylinders with an outer diameter of 2.5 cm are placed on the turntable as imaging targets, as shown in Fig. 8(a). Different from that in Fig. 7(a), the distance between adjacent cylinders is changed to 10 cm. Figures 8(b)-8(f) show the ISAR imaging results when the center frequencies of the LFM radar signal are 14, 20, 26, 32, and 38 GHz. As can be seen, the three cylinders can be clearly distinguished. The range resolution of the ISAR imaging is determined by the bandwidth of the radar signal, which is 3.75 cm. The cross-range resolution is determined by the center frequency of the radar signal, the accumulation time, and the target rotation angular velocity. With the increase of the center frequency of the radar signal, the cross-range resolution of the ISAR imaging improves from 3.91 to 1.03 cm. The improvement can be clearly observed by comparing Figs. 8(b)-8(f).

3.4 Spectrum sensing

The spectrum sensing function of the prototype is then tested. In this test, the center frequency of the IF LFM is set to 4 GHz, and the instantaneous analysis bandwidth is 2 GHz, which is determined by the bandwidth of the –2nd-order optical sideband. Furthermore, the frequency measurement range is adjusted by tuning the pump wave frequency by changing the frequency of the signal output from the MSG. Specifically, as the MSG signal frequency varies from 2.4 to 38.4 GHz in increments of 2 GHz, the corresponding frequency measurement range shifts accordingly, from 0 to 2 GHz initially to 36 to 38 GHz finally, also in increments of 2 GHz. An arbitrary waveform generator (AWG, Keysight M8195A) is used to generate a single-tone or two-tone RF SUT, while a synthesized sweeper (HP 83752B) is used to generate a frequency-sweep SUT with a period of 100 ms. An SUT transmitting antenna is connected to the two signal sources to radiate the SUT to free space, and $RX_2$ of the prototype collects the SUT in free space and the frequency information of the SUT is further analyzed by the prototype. It should be noted that since the SUT transmitting antenna can only work in the frequency range from 8 to 40 GHz, the AWG and synthesized sweeper are directly connected to the prototype by RF cables when the SUT frequency is within 8 GHz. In this case, the synthesized sweeper output power is set to –50 dBm, and the AWG output power is adjusted to around –45 dBm by attenuators. When the SUT frequency is greater than 8 GHz, the AWG and synthesized sweeper output is connected to the SUT transmitting antenna, and $RX_2$ is used to receive the SUT from the SUT transmitting antenna 1 m away, as shown in Fig. 9(a). In this case, the AWG and synthesized sweeper output power are set to 4 and 7 dBm, respectively. Furthermore, when $RX_2$ is used for spectrum sensing, the interference from the radar-transmitting antenna should be isolated to avoid self-interference. In this case, radar and spectrum sensing functions can be performed simultaneously.

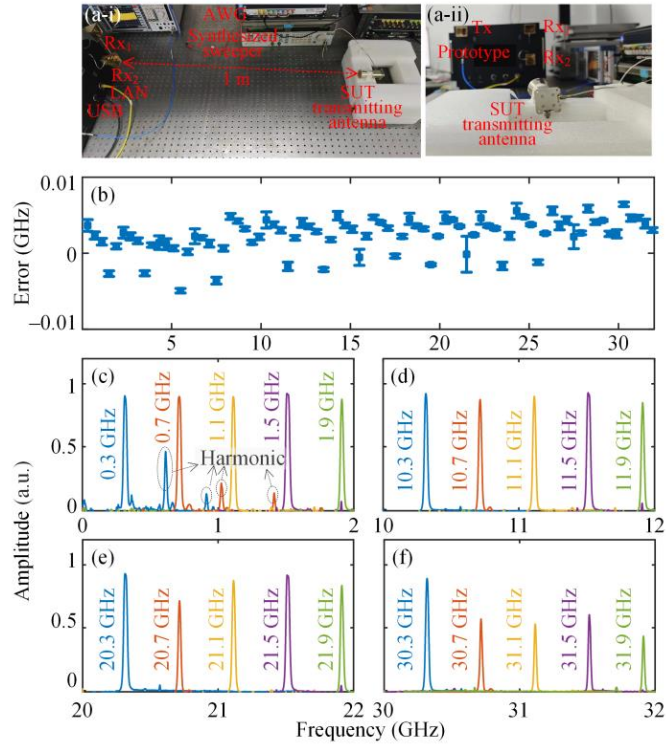

Fig. 9. (a) Photograph of the spectrum sensing setup. (b) Single-frequency measurement errors at different frequencies. Single-frequency measurement results when the frequency measurement range is (c) 0 to 2 GHz, (d) 10 to 12 GHz, (e) 10 to 12 GHz, and (f) 30 to 32 GHz.

Figures 9(b)-(f) show the frequency measurement results of the single-tone SUT. The error standard deviations of 10 measurements when the SUT frequency changes from 0.3 to 31.9 GHz with a step of 0.4 GHz are shown in Fig. 9(b). It can be seen that the measurement error is less than ±7 MHz. The temporal waveforms of the electrical pulses in a single sweep period corresponding to one measurement at different frequency bands are shown in Figs. 9(c)-(f). Since the maximum signal frequency that can be generated in our lab does not exceed 32 GHz using the AWG, only measurements up to 31.9 GHz are shown in Fig. 9. It should also be noted that the frequency of the signal can be calculated according to the generated electrical pulses by $2t_{sut}/T$ GHz $+ f_a + 2f_C - 10.4$ GHz, where $t_{sut}$ is the time instant corresponding to the peak value of the pulse within a measurement period $T$. In Fig. 9, the time axis is replaced with the corresponding frequency axis. It can be seen that different SUT frequencies correspond to pulses distributed at different locations. In Fig. 9(c), when the SUT frequency is 0.3 and 0.7 GHz, there are some extra frequency components, which are the harmonics of the signal generated by the AWG. In Fig. 9(f), the waveform amplitude gets smaller when the SUT frequency is larger than 30.3 GHz, which is mainly attributed to the limited bandwidth of the AWG. By accurately adjusting the pump wave frequency, the frequency measurement range can be broadened to span from 0 to 40 GHz, provided that the center frequency of the IF LFM signal exceeds 4.2 GHz. Conversely, when the center frequency of the IF LFM signal is 2.5 GHz, the minimum frequency measurement range is limited to 0 to 36.6 GHz.

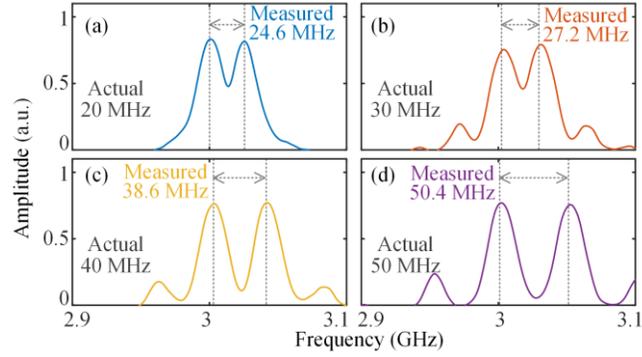

Fig. 10. Frequency measurement of a two-tone SUT when the frequency interval is (a) 20, (b) 30, (c) 40, and (d) 50 MHz.

Figure 10 shows the frequency measurement results of a two-tone test. The two-tone SUT has a fixed frequency at 3 GHz and another frequency tuned from 3.02 to 3.05 GHz with a step of 10 MHz, corresponding to a frequency interval from 20 to 50 MHz. It can be seen that the two-tone SUT can be distinguished even if the frequency interval is only 20 MHz, which means the frequency resolution of the frequency measurement function of the prototype is better than 20 MHz.

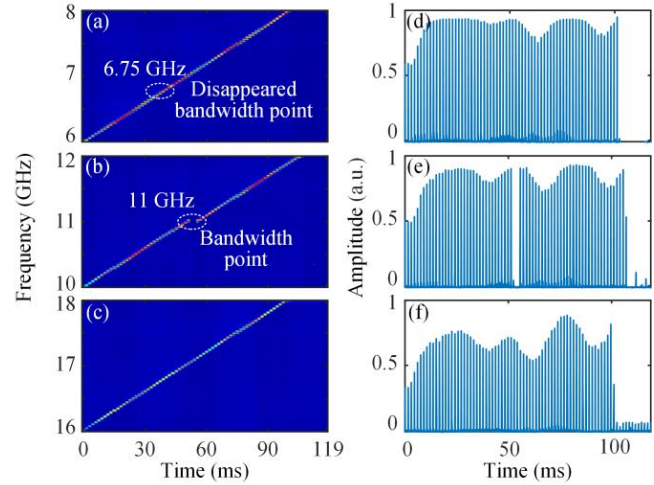

Fig. 11. Time-frequency analysis results of a frequency-sweep signal from (a) 6 to 8 GHz, (b) 10 to 12 GHz, and (c) 16 to 18 GHz. (d)-(f) are the temporal waveforms corresponding to (a)-(c).

As demonstrated in the above experiment, the frequency of a stationary signal can be identified in a single frequency-sweep period, i.e., 1.4876 ms in the prototype. Nevertheless, a non-stationary signal can still be analyzed using the prototype if it exhibits local stationarity over the 1.4876 ms frequency-sweep period. In this case, a time-frequency diagram is commonly used to characterize the SUT by accumulating the results of multiple frequency-sweep periods. Figures 11(a)-(c) show the time-frequency analysis results of a 100-ms frequency-sweep SUT generated by the synthesized sweeper when the frequency-sweep range of the SUT is 6 to 8 GHz, 10 to 12 GHz, and 16 to 18 GHz, respectively. The corresponding temporal waveforms are shown in Figs. 11(d)-(f). As can be seen, the linearly frequency-sweep characteristics of the SUT are well characterized by the time-frequency diagrams. However, in Fig. 11(b), a noticeable detail is that there is a discontinuity in the time-frequency diagram at 11 GHz, which is attributed to the fact that 11 GHz is a bandwidth point of the synthesized sweeper. A switch event at this point creates a sweep gap when the sweeper sweeps over the bandwidth point [40]. It is worth noting that 6.75 GHz is also a bandwidth point, but it is not visible in Fig. 11 (a) because the synthesizer switch is deactivated if the sweep range is less than 80% of the sweep starting frequency within the 2 to 11 GHz frequency band [40]. Additionally, the vacancy after the frequency-sweep SUT is the result of the retrace of the synthesized sweeper when a sweep is finished. Moreover, the waveform amplitudes in Figs. 11(d)-(f) are not uniform but have a similar shape, which is due to the use of the same –2nd-order LFM optical sideband in the test, and the LFM optical sideband is inherently not completely flat. The envelope of the waveform in Figs. 11(d)-(f) exhibit shapes similar to the radar signal spectrum at a 16 GHz center frequency, as illustrated in Fig. 3, because the same IF LFM signal center frequency is employed. Besides, by further reducing the sweep period $T$ of the IF LFM signal, the time required for a single measurement can be further decreased, enabling faster signal measurement. This will aid the system in measuring signals with more rapid frequency variations. It is worth noting that in the prototype, we do not further reduce the period, primarily due to limitations of the DDS in the IF LFM signal source. Further reduction in the period would significantly degrade the signal quality. Therefore, a high-quality and high-speed electrical frequency-sweep source is crucial for enhancing the performance of this system.

### 3.5 Comparison

A comparison between the proposed prototype and the previously reported photonics-assisted joint radar and spectrum sensing systems is given in Table 1. As can be seen, the advantages and key significance of this work are as follows: (1) To the best of our knowledge, this is the first reported microwave photonic prototype for joint radar and spectrum sensing applications; (2) This is the first joint radar and spectrum sensing system that can simultaneously implement all the functions shown in Table 1, including ranging, radial velocity measurement, ISAR imaging, frequency measurement, and time-frequency analysis; (3) Most of the key performance indicators of the prototype is better than the reported photonics-assisted joint radar and spectrum sensing systems.

**Table 1. Comparison of Different Photonics-Assisted Joint Radar and Spectrum Sensing Systems**

|  | Radar Frequency /Bandwidth | Ranging Resolution/ Error | Radial Velocity Error | ISAR Imaging Resolution | Frequency Measurement Range /Error/Resolution | Time-Frequency Analysis Ability /Prototype |
|---|---|---|---|---|---|---|
| Ref. [36] | 12-18 GHz /6 GHz | 2.6 cm/- | - | 2.6 cm × 2.8 cm | 28-37 GHz /±15 MHz/40 MHz | No/No |
| Ref. [37] | 18-26 GHz /8 GHz | 2.06 cm/- | - | - | 28-36 GHz /±16 MHz /37.6 MHz | No/No |
| Ref. [38] | 6-10 GHz /4 GHz | 4 cm/2.58 cm | 3.08 cm/s | - | 1-20 GHz /34.22 MHz/40 MHz | No/No |
| Ref. [39] | 12-18 GHz /6 GHz | 1.25 cm/- | - | 1.25 cm × 1.33 cm | 0.05-39.95 GHz /±50 MHz/20 MHz | No/No |
| This work | 8-40 GHz /4 GHz | 3.75 cm/±2 cm | ±1 cm/s | 3.75 cm × 1.03 cm [a] | 0-40 GHz /±7 MHz/<20 MHz [b] | Yes/Yes |

[a] The best cross-range resolution is 1.03 cm at a radar center frequency of 38 GHz.
[b] 20 MHz is the minimum frequency interval measured in the experiment, and the resolution can be far better than 20 MHz from the results in Fig. 10.

### 3.6 Application of the prototype in cognitive radio scenarios.

The proposed prototype can be applied to cognitive radio systems, intelligent transportation systems, and electronic warfare systems. Fig. 12 shows the schematic diagram of the prototype's application in cognitive radio scenarios. In cognitive radio systems, the spectrum sensing function of the prototype enables rapid scanning of the electromagnetic spectrum within the range of 0 to 40 GHz. Therefore, the presence of interferences and already utilized frequency bands can be quickly identified and positioned, and the spectrum holes can be acquired. According to spectrum holes found via the spectrum sensing function, the prototype can quickly adjust the radar signal frequency and set it within the spectrum holes, thus avoiding interference to the radar function and impacts from other signals. It is worth noting that since the radar and spectrum sensing functions can operate independently and simultaneously within 40 GHz, even when the radar system is in operation, the spectrum sensing function can continue to scan the entire electromagnetic spectrum continuously. Once any interference is detected within the radar's operating frequency band, the prototype can once again perform frequency agility on the radar signal, enabling it to operate normally without being disrupted.

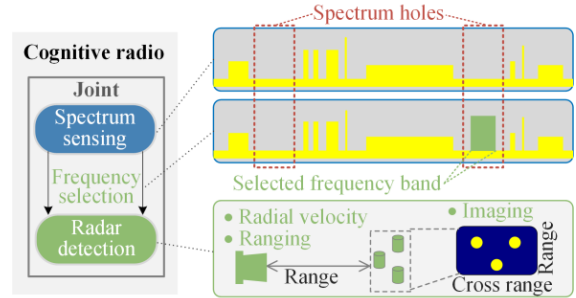

Fig. 12. Schematic diagram of the prototype's application in cognitive radio scenarios.

## 4. Conclusion

In summary, we have proposed, designed, built, and investigated a microwave photonic prototype for concurrent radar detection and spectrum sensing. At the system level, through the microwave photonic design, we achieve true integration by sharing hardware and signals between radar and spectrum sensing. Functionally, both radar and spectrum sensing can operate concurrently and be tuned independently within the vast 8-40 GHz bandwidth, which cannot be achieved before. In this prototype, the radar signal has an instantaneous bandwidth of 4 GHz and a tunable frequency range from 8 to 40 GHz, whereas the spectrum sensing instantaneous bandwidth is 2 GHz and the spectrum sensing frequency range can be tuned from 0 to 40 GHz. Moreover, compared to previous reports, this is the first prototype-level integration of microwave photonic radar and spectrum sensing, surpassing previous desktop experiments. Additionally, the prototype is also the first microwave photonic joint radar and spectrum sensing system with simultaneous radar ranging, velocity measurement, imaging, frequency measurement, and time-frequency analysis capabilities. A comprehensive investigation of the prototype is carried out. A ranging error of less than ±2 cm, a radial velocity error of less than ±1 cm/s, clear imaging of multiple small targets, a frequency measurement error of less than ±7 MHz, and a frequency resolution of better than 20 MHz can be achieved. The prototype is expected to be applied to cognitive radio systems, intelligent transportation systems, and electronic warfare systems after further miniaturization and integration design.

## Acknowledgements
This work was supported by the National Natural Science Foundation of China under Grant 62371191, the Space Optoelectronic Measurement and Perception Laboratory, Beijing Institute of Control Engineering under Grant LabSOMP-2023-05, and the Science and Technology Commission of Shanghai Municipality under Grant 22DZ2229004.